\documentclass[journal]{IEEEtran}
\usepackage{amsmath}
\usepackage{amsfonts}
\usepackage{epsfig}
\usepackage{amssymb}
\usepackage{graphics}
\usepackage{}
\usepackage{float}
\usepackage{amsmath}
\usepackage{slashbox}
\usepackage[T1]{fontenc}
\usepackage{cite}
\usepackage{cleveref}
\usepackage{array}
\usepackage{caption}
\usepackage{subcaption}
\usepackage[usenames, dvipsnames]{color}

\makeatletter
\newcommand{\thickhline}{%
    \noalign {\ifnum 0=`}\fi \hrule height 1.2pt
    \futurelet \reserved@a \@xhline
}
\newcolumntype{"}{@{\hskip\tabcolsep\vrule width 1pt\hskip\tabcolsep}}
\makeatother

\begin{document}

\title{The Max-Product Algorithm Viewed as Linear Data-Fusion: A Distributed Detection Scenario}

\author{\authorblockN{Younes~Abdi,~\IEEEmembership{Member,~IEEE,} and Tapani~Ristaniemi,~\IEEEmembership{Senior Member,~IEEE}}
\thanks{Y.~Abdi and T.~Ristaniemi are with the Faculty of Information Technology, University of Jyv\"askyl\"a, P.~O.~Box 35, FIN-40014, Jyv\"askyl\"a, Finland, Tel. +358 40 7214 218 \mbox{(e-mail:younes.abdi@jyu.fi, tapani.ristaniemi@jyu.fi)}. 
}\vspace{-0.25 in}
}

\maketitle

\begin{abstract}
In this paper, we disclose the statistical behavior of the max-product algorithm  configured to solve a \emph{maximum a posteriori} (MAP) estimation problem in a network of distributed agents. Specifically, we first build a distributed hypothesis test conducted by a max-product iteration over a binary-valued pairwise Markov random field and show that the decision variables obtained are linear combinations of the local log-likelihood ratios observed in the network. Then, we use these linear combinations to formulate the system performance in terms of the false-alarm and detection probabilities. Our findings indicate that, in the hypothesis test concerned, the optimal performance of the max-product algorithm is obtained by an optimal linear data-fusion scheme and the behavior of the max-product algorithm is very similar to the behavior of the sum-product algorithm. Consequently, we demonstrate that the optimal performance of the max-product iteration is closely achieved via a linear version of the sum-product algorithm which is optimized based on statistics received at each node from its one-hop neighbors. Finally, we verify our observations via computer simulations. 
\end{abstract}

\begin{keywords}
Statistical inference, distributed systems, max-product algorithm, sum-product algorithm, linear data-fusion, Markov random fields, factor graphs, spectrum sensing. 
\end{keywords}

\section{Introduction}\label{sec:Intro}
\PARstart{S}{tandard} optimization methods are computationally demanding when dealing with a large collection of correlated random variables. This is a well-known challenge in designing distributed statistical inference techniques  used in a wide range of signal-processing applications such as channel decoding, image processing, spread-spectrum communications, distributed detection, etc. Alternatively, message-passing algorithms over factor graphs  provide a powerful low-complexity approach to characterizing and optimizing the collective impact of those variables on the desired system performance, see e.g., \cite{Wainwright08, Loeliger04, Cetin06}. Consequently, a better understating of the statistical behavior of the message-passing algorithms leads to statistical inference systems with better performance. Two widely-used message-passing algorithms are the so-called sum-product and max-product algorithms. We have analyzed the behavior of the sum-product algorithm, a.k.a., the belief propagation algorithm, in \cite{Abdi20}. 

 Our main focus in this paper is on the max-product algorithm which is an iterative method for approximately solving the problem of \emph{maximum a posteriori} (MAP) estimation \cite{Wainwright04}. We analyze the behavior of this algorithm in a distributed detection scenario where every network node estimates a binary-valued random variable based on noisy observations collected throughout the entire network. The correlations between the random variables are modeled by a pairwise Markov random field (MRF) \cite{Wainwright08} whose structure fits well into pairwise interactions between the nodes in an ad-hoc network configuration. An MRF is an undirected graph where vertices correspond to the random variables of interest and edges represent the  correlations between them. By using the max-product algorithm, the estimation problem concerned is decomposed into a number of small optimizations performed locally at each node based on information provided by other nodes in the network via one-hop communications per iteration. 
 
 \subsection{Max-Product v.s Sum-Product} 
Let $\boldsymbol{x} = [x_1, ..., x_N]^T$ denote a vector of $N$ discrete-valued random variables to be estimated given the observations $\boldsymbol{Y} \triangleq [\boldsymbol{y}_1, ..., \boldsymbol{y}_N]$ where $\boldsymbol{y}_i \triangleq [y_i(1), ..., y_i(K)]^T$ denotes $K$ samples collected at node $i$ and $i = 1, ..., N$. The MAP estimation of $\boldsymbol{x}$ is formally stated as 
\begin{equation} \label{mapEq}
\hat{\boldsymbol{x}} = \arg \max_{\boldsymbol{x}} p(\boldsymbol{x} \vert \boldsymbol{Y}).
\end{equation}
 This is an integer program, which is NP-hard. When $p(\boldsymbol{x} \vert \boldsymbol{Y})$ is stated in the form of an MRF, \eqref{mapEq} can be solved with low complexity by using two commonly-used message-passing algorithms, i.e., the sum-product and max-product algorithms. We provide a brief overview of the origins and differences of the two algorithms here. An interested reader may refer to \cite{Wainwright08, Loeliger04, Cetin06, Wainwright04} for further details and more comprehensive discussions. 
 
 The optimization in \eqref{mapEq} can be approximated by the so-called \emph{Bethe variational problem} \cite[Sec. 4]{Wainwright08} and also by a \emph{linear program} \cite[Sec. 8]{Wainwright08}. For a tree-structured MRF, both of these approximations turn out to be convex, can be solved by the Lagrangian dual method, and provide exact solutions for \eqref{mapEq}. The sum-product iteration solves the dual of the Bethe problem while the max-product algorithm solves the dual of the linear program. When the MRF contains cycles, the fixed points of both message-passing algorithms provide approximate solutions for the MAP estimation problem concerned.  

The sum-product algorithm solves \eqref{mapEq} by finding the marginal distributions associated with $p(\boldsymbol{x} \vert \boldsymbol{Y})$ whereas the max-product is an attempt to find the so-called max-marginals. More specifically, the sum-product algorithm gives the marginal distribution, at node $i$, defined as 
\begin{equation} \label{pmarginal}
p(x_i \vert \boldsymbol{Y}) \triangleq \sum_{\{\boldsymbol{x}' \vert x'_i = x_i\}}p(\boldsymbol{x}' \vert \boldsymbol{Y})
\end{equation} 
that is used to solve \eqref{mapEq} by 
\begin{equation}\label{argmaxp}
\hat{x}_i = \arg \max_{x_i} p(x_j \vert \boldsymbol{Y}).
\end{equation}
The outcome of the max-product algorithm at node $i$ is the max-marginal distribution of $x_i$ defined as 
\begin{equation} \label{maxMarginal}
q(x_i \vert \boldsymbol{Y}) \triangleq \kappa \max_{\{\boldsymbol{x}' \vert x'_i = x_i\}} p(\boldsymbol{x}' \vert \boldsymbol{Y}),
\end{equation}
where $\kappa$ denotes a positive arbitrary normalization constant. If for each node, the maximum of $q(x_i \vert \boldsymbol{Y})$ is attained at a unique value, then the MAP configuration is unique and can be obtained by maximizing the corresponding max-marginal at each node \cite{Wainwright04}, i.e.,
\begin{equation} \label{eq:maxMaxMargin}
\hat{x}_i = \arg \max_{x_i} q(x_i \vert \boldsymbol{Y}).
\end{equation}
In case there is a node at which the maximum of $q(x_i \vert \boldsymbol{Y})$ is not attained at a unique value, Eq. \eqref{eq:maxMaxMargin} provides a sub-optimal solution. We discuss this case in Section \ref{subsec:Discussion}.
 The analysis proposed in this paper along with the work in \cite{Abdi20} shows that, in a distributed detection scenario, both methods are equivalent to a linear data-fusion. 

\subsection{Related Work and Motivation}
Many existing works on the max-product algorithm attempt to pave the way towards theoretical guarantees on the convergence of the algorithm and on the quality of the resulting fixed points on graphs with arbitrary topology and with arbitrary probability distributions, see e.g., \cite{Wainwright04, Weiss01, Kolmogorov05, Wainwright05}. This is still an open and growing research field. There also exist numerous works that tailor the max-product iteration into a particular statistical inference scenario taking into account the graph structure and available resources in that particular setting. Examples of such works can be found in LDPC decoding \cite{Zhao05}, multi-sensor target tracking \cite{Chen03, Chen06}, clock synchronization in WSNs \cite{Ahmad12, Zennaro13}, sparse code multiple access \cite{Han19}, etc. Moreover, several works in the literature use some sort of approximation in modeling various message-passing structures to offer a deeper insight into the behavior of message-passing algorithms or to propose better distributed inference methods, see e.g., \cite{Bayati11, Donoho09, Donoho10, Rangan2010, Rangan2011}. 

To the best of our knowledge, the existing works do not offer a comprehensive analysis and optimization framework for the max-product algorithm in the context of distributed detection. Formulating the performance of a wireless sensor network that employs distributed detection requires understanding the statistical behavior of the underlying data-exchange process between the sensing nodes. When the max-product algorithm is used, this data-exchange process is built based on the structure of the factor graph that models the network behavior. Therefore, an optimal system design calls for finding the relation between the parameters of the factor graph and the network performance metrics. More specifically, an optimal design requires answering the following questions: 
\begin{itemize}
  \item How to best represent the network behavior by a pairwise MRF and how to impose certain constraints on the system performance, in terms of the desired  false-alarm or detection probabilities, when the data-exchange process between the nodes is realized by the max-product algorithm over that MRF? 
 \end{itemize}
We answer these questions in this paper. The importance of the research gap discussed here is highlighted by noting that distributed detection is a major functionality in many advanced communication scenarios such as industrial internet of things, internet of vehicles, mobile crowdsensing, and cognitive radio (CR) networks, see e.g., \cite{Ganti11, Cao19, Sun17, Sun19, Penna12}.

\subsection{Contribution}
We show that the max-product algorithm works as a linear data-fusion process. Linear fusion schemes are commonly used in distributed detection systems to achieve near-optimal performance with low implementation complexity, see e.g., \cite{Quan08, Quan10, Abdi14, Abdi17}. Therefore, we indicate that the knowledge already developed in linear distributed detection methods can be used to better understand the behavior of the max-product algorithm. The proposed analysis is supported by a strong connection between the sum-product and max-product operations. In particular, we show that, in the distributed detection scenario concerned, the behavior of the max-product algorithm is very similar to the behavior of the sum-product algorithm and that the decision variables built by the max-product operation are linear combinations of the local likelihoods in the network---a behavior we have already observed in the sum-product algorithm \cite{Abdi20}. By using this linearity, we make the following contributions:
\begin{itemize}
  \item We show that the message-update rule in the max-product algorithm is almost the same as its counterpart in the sum-product algorithm. 
  \item We show that when performing a distributed MAP estimation via the max-product algorithm over a network modeled by a pairwise MRF, under certain practical conditions, the decision variables obtained are linear combinations of the local log-likelihood ratios (LLR) in the network. 
 \item We find the probability distribution function of the decision variables in a practical detection scenario and formulate the detection performance in closed form.  
 \item We show how to set the detection threshold to achieve a predefined detection performance. 
 \item We show that the optimal linear message-passing algorithm in \cite{Abdi20} attains the optimal detection performance of the max-product algorithm in the distributed detection scenario concerned.  
\end{itemize}

As in \cite{Penna12, Wymeersch12}, and \cite{Abdi20}, we clarify our findings by considering a spectrum sensing scheme in a CR network. In these networks, the wireless nodes perform spectrum sensing, in bands allocated to the so-called primary users (PU), in order to discover vacant parts of the radio spectrum and establish communication on those temporarily- or spatially-available spectral opportunities \cite{Akyildiz11}. In this context, CRs are considered secondary users (SU) in the sense that they have to vacate the spectrum, to avoid making any harmful interference, once the PUs are active.

The rest of the paper is organized as follows. In Section \ref{section:MAP}, we formulate the MAP estimation problem and discuss how to solve it in a network of distributed agents via the sum-product and max-product algorithms. In addition, we illustrate in Section \ref{section:MAP} the connection between the sum-product and max-product operations. Then, we analyze the behavior of the max-product algorithm in Section \ref{sec:Analysis} to show that it works as a linear fusion scheme. In Section \ref{sec:linearFusion}, we briefly discuss the use of linear data-fusion in distributed detection along with the proposed optimization framework. We then verify our analysis by computer simulations in Section \ref{sec:Simulations} and present our concluding remarks in Section \ref{sec:conclusion}.

 \section{Distributed Detection via Message-Passing}\label{section:MAP}
\begin{figure}[]
	\centering 
  \includegraphics[scale=0.28]{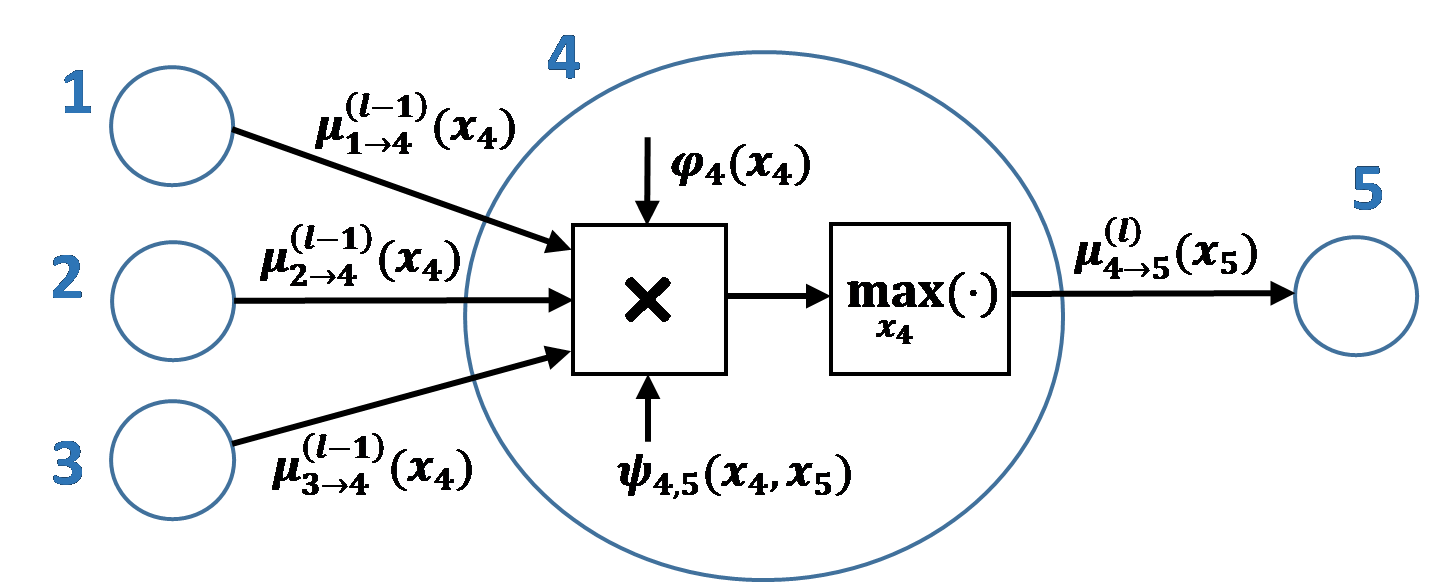} 
  \caption{An schematic diagram of the max-product algorithm illustrating how the messages are generated. }
  \label{fig:maxProd}
\end{figure}

We consider a pairwise MRF defined on an undirected graph $G = (\mathcal{V,E})$ composed of a set of $N$ vertices or nodes $\mathcal{V} \triangleq \{1, ..., N\}$ and a set of edges $\mathcal{E} \subset \mathcal{V} \times \mathcal{V}$. Each node $i \in \mathcal{V}$ corresponds to a random variable $x_i$ and each edge $(i,j) \in \mathcal{E}$, which connects nodes $i$ and $j$, represents a possible correlation between random variables $x_i$ and $x_j$.  The MRF is used to factorize the a posteriori distribution function $p({\boldsymbol x}\vert {\boldsymbol Y}) $ into single-variable and pairwise terms, i.e.,    
\begin{equation}\label{eq:MRF} 
p({\boldsymbol x}\vert {\boldsymbol Y}) \propto  \prod_{n \in \mathcal{V}} \phi_{n}(x_{n}) \prod_{(i,j) \in {\cal E}} \psi_{ij}(x_i, x_j),
\end{equation}
where $\propto$ denotes proportionality up to a multiplicative constant. 
In our detection scenario, the main goal of each node, say node $i$, is to find its max-marginal a posteriori distribution $q(x_i \vert \mathbf{y})$. This goal is achieved by the max-product algorithm where the messages sent from node $k$ to node $j$ in the network are built as  
\begin{align}\label{eq:mu_kj}
\mu_{k \to j}^{(l)}(x_j) &\propto \max_{x_k}  \Bigg [ \phi_k(x_k)  \psi_{kj}(x_k,x_j) 
\prod_{n \in \mathcal{N}_k^j}\mu_{n \to k}^{(l-1)}(x_k)  \Bigg  ],
\end{align}
where $\mathcal{N}_k^j$ denotes the set of neighbors of node $k$ except for node $j$. Fig. \ref{fig:maxProd} illustrates this process. The \emph{belief} of node $j$ at iteration $l$, denoted $b_j^{(l)}(x_j)$, is formed by multiplying the local inference result $\phi_{j}(x_j)$ by all the messages received from its neighbors, i.e., $b_j^{(l)}(x_j) \propto \phi_{j}(x_j) \prod_{k \in \mathcal{N}_j} \mu_{k \rightarrow j}^{(l)}(x_j) $. The resulting belief is then used to estimate the desired max-marginal distribution, i.e., $ q(x_j \vert \boldsymbol{Y}) \approx b_j^{(l)}(x_j)$. We can express $b_j^{(l)}(x_j)$ in the logarithm form as
\begin{equation}\label{eq:ln_b} 
\ln b_j^{(l)}(x_j) = \ln \phi_{j}(x_j) + \sum_{k \in \mathcal{N}_j}  m_{k \rightarrow j}^{(l)} (x_j),
\end{equation}
where 
\begin{align}\label{eq:m_kj} 
&m_{k \to j}^{(l)}(x_j)  \triangleq \ln \mu_{k \to j}^{(l)}(x_j)
 \nonumber \\
&= \max_{x_k}  \Bigg  [ \ln \phi_k(x_k) + \ln \psi_{kj}(x_k,x_j)  
  + \sum_{n \in \mathcal{N}_k^j} m_{n \to k}^{(l-1)}(x_k) \Bigg ].
\end{align}
We have replaced $\propto$ by equality in our formulations since the proportionality constant turns into an offset value in the log domain with no impact on the proposed analysis. We adopt the commonly-used exponential model to represent the probability measure defined on $\boldsymbol{x}$, i.e., 
 \begin{equation}\label{eq:p(x)} 
p({\boldsymbol x}) \propto  \textup{exp}\left ( \sum_{n \in \mathcal V} \theta_{n}x_{n} + \sum_{(i,j) \in {\cal E}} J_{ij}x_i x_j \right ).
\end{equation}
Consequently, from $p({\boldsymbol x}\vert {\boldsymbol Y}) = {p({\boldsymbol Y}\vert {\boldsymbol x})p({\boldsymbol x})/p({\boldsymbol Y})}$, we obtain \cite{Abdi20}
\begin{equation}\label{eq:p(x|y)_2} 
p({\boldsymbol x}\vert {\boldsymbol Y}) \propto \prod_{k \in \mathcal{V}} p(\boldsymbol{y}_k \vert x_k)  \prod_{(i,j) \in \mathcal{E} } e^{J_{ij}x_ix_j},
\end{equation}
where the proportionality sign covers $1 \over p({\boldsymbol Y})$. Since $\theta_k$ does not affect the proposed analysis, we have set $\theta_k = 0$ for all $k$. By comparing \eqref{eq:p(x|y)_2} to \eqref{eq:MRF}, we obtain 
\begin{eqnarray}
 \phi_k(x_k) \triangleq p(\boldsymbol{y}_k \vert x_k) \label{eq:phi}, \\
 \psi_{kj}(x_k,x_j) \triangleq e^{J_{kj}x_kx_j} \label{eq:psi}.
 \end{eqnarray}
Assuming Gaussian observations at the nodes, we have 
\begin{equation} \label{eq:p_gauss}
p(\boldsymbol{y}_k \vert x_k) = \frac{1}{\sqrt{2\pi}\sigma}\exp\left ( \frac{-1}{2\sigma^2}\left \| \boldsymbol{y}_k - \boldsymbol{\mu}_{\boldsymbol{y}_k | x_k}  \right \|^2 \right ),
\end{equation}
where $\boldsymbol{\mu}_{\boldsymbol{y}_k |x_k} \triangleq E[\boldsymbol{y}_k |x_k]$. For $i=1, ..., K$, we have $y_k(i)  = \xi_k s_k(i) + \nu_k(i)$ where $\xi_k \triangleq \frac{1}{2}(x_k+1)$ and $\nu_k(i) \sim \mathcal{N}(0,\sigma^2)$. 
In the vector format, we have $\boldsymbol{y}_k = \xi_k \boldsymbol{s}_k  + \boldsymbol{\nu}_k$ where $\boldsymbol{s}_k \triangleq [s_k(1), ..., s_k(K)]^T$ denotes a deterministic but unknown sequence of PU signal samples received at node $k$ and $\boldsymbol{\nu}_k \triangleq [\nu_k(1), ..., \nu_k(K)]^T$. Hence,  $\boldsymbol{\mu}_{\boldsymbol{y}_k |x_k} = \xi_k \boldsymbol{s}_{k}$. We have $x_k \in \{-1, +1\}$ for all $k \in \mathcal{V}$ while $\xi_k \in \{0,1\}$ maps the state of $x_k$ to the occupancy state of the radio spectrum sensed by node $k$. See Table I for a list of symbols used in this paper.

\begin{table}[] 
\normalsize
\caption{Main Parameters Specifying the Detector Structure}
\label{tab:params}
\begin{footnotesize}  
\begin{tabular}{ l l }
\hline 
\textbf{Symbol} & \textbf{Meaning}   \\
\hline \hline
$N$ & Number of sensing nodes in the network\\
$K$ & Number of samples collected at each sensing node\\
$\mathcal{V}$ & Set of vertices in the factor graph\\
$\mathcal{E}$ & Set of edges in the factor graph\\
$\mathcal N_j$ & Set of neighbors of node $j$\\
$\mathcal N_j^k$ & Set of neighbors of node $j$ except for node $k$\\
$\mu_{k \to j}^{(l)}(x_j)$ & Max-product message sent to node $j$ from node $k$ \\
 $ \bar \mu_{k \to j}^{(l)}(x_j)$ & Sum-product message sent to node $j$ from node $k$ \\
$b_{j}^{(l)}(x_j)$ & Max-product beliefs \\
$\bar b_{j}^{(l)}(x_j)$  & Sum-product beliefs \\
$m_{k \to j}^{(l)}(x_j)$  & Max-product message in the log domain \\
$\bar m_{k \to j}^{(l)}(x_j)$ & Sum-product message in the log domain \\
$\lambda_j^{(l)}$ & Max-product decision variable at node $j$ \\
$\bar \lambda_j^{(l)}$ & Sum-product decision variable at node $j$ \\
$\delta_{k \to j}^{(l)}$  & LLR of max-product messages \\
$\bar \delta_{k \to j}^{(l)}$  & LLR of sum-product messages \\
$\tau_j$ &  Max-product detection threshold  \\
$\bar \tau_j$ &  Sum-product detection threshold  \\
$\gamma_j$ &  Local sensing outcome at node $j$ \\
$E_j$ & Energy of the PU signal received at node $j$ \\
$\boldsymbol{s}_j$ & Signal to be detected at node $j$\\
$\boldsymbol{y}_j$ & Noisy received signal at node $j$\\
$\boldsymbol{\nu}_j$ & Noise received at node $j$\\
$\boldsymbol{\mu}_{\boldsymbol{y}_j |x_j}$ & Conditional mean of $\boldsymbol{y}_j$ given $x_j$\\
$\sigma^2$ & Noise variance\\
$\xi_j$ & Auxiliary variable that maps $x_j$ to $\{0,1\}$\\
$f(\boldsymbol{x})$ & Transfer function built  by the sum-product operation\\
$P_{\textup{f}}^{(j)}$ & False-alarm probability at node $j$ \\
$P_{\textup{d}}^{(j)}$ & Detection probability at node $j$ \\
$\phi_{n}(x_{n})$ & Single-variable factor in $p({\boldsymbol x}\vert {\boldsymbol Y})$ \\
$\psi_{ij}(x_i, x_j) $ & Pairwise factor in $p({\boldsymbol x}\vert {\boldsymbol Y})$ \\
$\theta_n$ & Single-variable exponent factor in $p(\boldsymbol x)$\\
$J_{ij}$ & Pairwise exponent factor in $p(\boldsymbol x)$\\
$T$ & Sample-window size in updating $J_{ij}$'s \\
$\zeta$ &  Determines the impact of  $T$ samples on $J_{ij}$'s\\
$\hat{x}_k^{(l)}(x_j)$ &  Outcome of the MLE at node $k$, equals to $u_{kj}^{(l)} + v_{kj}^{(l)}x_j$ \\
$\hat{x}_k$ &  Outcome of the desired MAP estimation at node $k$ \\
\hline 
\end{tabular}
\end{footnotesize}
\vspace*{0pt}
\end{table} 

The max-marginals obtained by the max-product algorithm are used to conduct a distributed MAP estimation as in \eqref{eq:maxMaxMargin}. Specifically, after $l$ iterations, at each node the approximate max-LLR is built and compared, as a decision variable, to a predefined threshold, i.e.,  
\begin{equation}\label{eq:lambda_j} 
\lambda_j^{(l)} \triangleq \ln \frac{b_j^{(l)}(x_j = +1)}{b_j^{(l)}(x_j = -1)} \gtrless \tau_j,
\end{equation}
which means that $\hat{x}_j = +1$ if $\lambda_j^{(l)} > \tau_j$ and $\hat{x}_j = -1$ otherwise. Note that $\lambda_j^{(l)} \approx \ln \frac{q(x_j = +1 \vert \boldsymbol{Y})}{q(x_j = -1 \vert \boldsymbol{Y})}$. To see the impact of messages on the decision variable, we express $\lambda_j^{(l)}$ as 
\begin{equation}\label{eq:lambda_j+delta} 
\lambda_j^{(l)} = \gamma_j + \sum_{k \in \mathcal{N}_j} \delta_{k \to j}^{(l)},
\end{equation} 
where $\gamma_j$ denotes the local LLR obtained at node $j$, i.e., $\gamma_j \triangleq \ln \frac{\phi_j(x_j = +1)}{\phi_j(x_j = -1)} = \ln \frac{p(\boldsymbol{y}_j \vert x_j = +1)}{p(\boldsymbol{y}_j \vert x_j = -1)}$ while $\delta_{k \to j}^{(l)}$ denotes the LLR of the messages at iteration $l$, i.e., 
\begin{align} 
\delta_{k \to j}^{(l)} &\triangleq \ln \frac{\mu_{k \to j}^{(l)}(x_j = +1)}{\mu_{k \to j}^{(l)}(x_j = -1)} \nonumber \\
&= m_{k \to j}^{(l)}(x_j = +1) - m_{k \to j}^{(l)}(x_j = -1) \label{eq:delta_kj}.
\end{align}
By using the signal model in \eqref{eq:p_gauss}, we obtain
\begin{equation} \label{eq:matched}
\gamma_j = \boldsymbol{s}_j^T\boldsymbol{y}_j - {1 \over 2}E_j,
\end{equation}
where $E_j \triangleq \left \| \boldsymbol{s}_j\right \|^2$. Consequently, it is clear that, given $x_j$, the local LLR $\gamma_j$ follows a Gaussian distribution. For simplicity we assume that $\sigma^2 = 1$. Eq. \eqref{eq:matched} indicates a matched filtering process, a.k.a., coherent detection \cite{Ma09} performed locally at each sensing node. In practice, since $\boldsymbol{s}_k$ is unknown, energy detection is used as the local sensing scheme \cite{Abdi20, Penna12}. That is, the local sensing outcome is formed as 
\begin{equation} \label{eq:energyDet}
\gamma_j \triangleq {1 \over K}\left \| \boldsymbol{y}_j\right \|^2 - \tau_0,
\end{equation}
where $\tau_0$ is set such that $\textup{Pr}\{\gamma_k > 0 \vert x_k = 0\} = \alpha$, i.e., $\tau_0 = \sigma_\nu^2 \left (1+\sqrt{\frac{2}{K}}Q^{-1}(\alpha) \right )$ where $Q^{-1}(\cdot)$ denotes the inverse of the $Q$-function. Assuming the number of signal samples $K$ is large enough \cite{Quan08, Quan10, Abdi14}, the central limit theorem states that, given $x_j$, the sensor outcome $\gamma_j$ in \eqref{eq:energyDet} follows a Gaussian distribution.

The sum-product algorithm has a similar structure except that the max operator in \eqref{eq:m_kj} is replaced by a summation. This message-update rule is given by 
\begin{align}\label{eq:mu_kj_bar}
\bar{\mu}_{k \to j}^{(l)}(x_j) &\propto \sum_{x_k}  \Bigg [ \phi_k(x_k)  \psi_{kj}(x_k,x_j) 
\prod_{n \in \mathcal{N}_k^j}\bar{\mu}_{n \to k}^{(l-1)}(x_k)  \Bigg  ].
\end{align}
The beliefs made by the sum-product algorithm are denoted $\bar{b}_j(x_j)$ in this paper and calculated by \eqref{eq:ln_b}  in which $\mu_{k \to j}^{(l)}$ is replaced by  $\bar{\mu}_{k \to j}^{(l)}$. The sum-product algorithm approximates the marginal distributions of the random variables of interest, i.e., $\bar{b}_j(x_j) \approx p(x_j |\boldsymbol{Y}) $.  The detection process is conducted by comparing the resulting decision variable to a predefined threshold as in \eqref{eq:lambda_j} where $\lambda_j$, $b_j(x_j)$, and $\tau_j$ are replaced by $\bar \lambda_j$, $\bar{b}_j(x_j)$, and $\bar \tau_j$, respectively. 

Similar to \eqref{eq:lambda_j+delta}, the detection variable build by the sum-product iteration can be expressed as 
\begin{equation}\label{eq:lambdabar_j} 
\bar{\lambda}_j^{(l)} = \gamma_j + \sum_{k \in \mathcal{N}_j} \bar{\delta}_{k \to j}^{(l)},
\end{equation}
 where  $\bar{\delta}_{k \to j}^{(l)} \triangleq \bar{m}_{k \to j}^{(l)}(+1) - \bar{m}_{k \to j}^{(l)}(-1)$ while $\bar{m}_{k \to j}^{(l)}(x_j) \triangleq \ln \bar{\mu}_{k \to j}^{(l)}(x_j)$. Through some algebra, we obtain
\begin{equation}\label{EQ14} 
 \bar{\delta}_{k \to j}^{(l)} = S \left (J_{kj}, ~\gamma_k + \sum_{n \in \mathcal{N}_k^j} \bar{\delta}_{n \to k}^{(l-1)} \right ),
\end{equation}
where $S(a,b) \triangleq \ln \frac{1+e^{a+b}}{e^a + e^b}$. 
$J_{kj}$ is determined by a moving average of length $T$ time slots, i.e., 
\begin{equation}\label{lrnFact} 
J_{kj} \triangleq \frac{\zeta}{T}\sum_{t=1}^{T} \left [ \boldsymbol{1}\{\hat{x}_j(t) = \hat{x}_k(t)\}-\boldsymbol{1}\{\hat{x}_j(t) \neq \hat{x}_k(t)\} \right ],
\end{equation}
 where $\zeta$ is a constant and $\boldsymbol{1}\{\cdot\}$ denotes the indicator function.

 
 
We use the first-order Taylor series expansion of $S$ to linearize the message-update rule as 
\begin{equation} \label{eq:delta_sumprod}
\bar{\delta}_{k \to j}^{(l)} \approx c_{jk}\left (\gamma_k + \sum_{n \in \mathcal{N}_k^j} \bar{\delta}_{n \to k}^{(l-1)} \right ),
\end{equation}
where $c_{jk} = \frac{(e^{2J_{kj}}-1)}{(1+e^{J_{kj}})^2}$ \cite{Abdi20}. Consequently, at node $j$ we have 
\begin{align}\label{eq:lambda_j_linear} 
\bar{\lambda}_j  \approx \gamma_j &+ \sum_{k \in \mathcal{N}_j}c_{jk}\gamma_k + \sum_{k \in \mathcal{N}_j}\sum_{n \in \mathcal{N}_k^j}c_{jk}c_{kn} \gamma_n  \nonumber\\
  &+ \sum_{k \in \mathcal{N}_j}\sum_{n \in \mathcal{N}_k^j}\sum_{m \in \mathcal{N}_n^k}c_{jk}c_{kn}c_{nm} \gamma_m + ... ,
\end{align}
where $\bar{\lambda}_j \triangleq \lim_{l \to \infty} \bar{\lambda}_j^{(l)}$. Eq. \eqref{eq:lambda_j_linear} shows that the sum-product algorithm is approximately a linear fusion scheme. Since the local LLRs are normal random variables, given the state of $x_i$'s, the decision variable $\bar{\lambda}_j$ is, approximately, a normal random variable as well. We can express this linear fusion in compact form as
 $\bar{\lambda}_j \approx \sum_{k \in \mathcal{V}} \bar{w}_{jk}\gamma_k$ where $\bar{w}_{jk}$ denotes the weight of $\gamma_k$ in this combination. 
  
To see the connection between the sum-product and max-product operations, let us take a closer look at the messages in the sum-product algorithm. In the log domain, we have
\begin{align} \label{eq:m_kj_bar} 
&\bar{m}_{k \to j}^{(l)}(x_j) \triangleq \ln \bar{\mu}_{k \to j}^{(l)}(x_j) =
\nonumber \\
& \ln \sum_{x_k} \exp  \Bigg  [ \ln \phi_k(x_k) + \ln \psi_{kj}(x_k,x_j) + \sum_{n \in \mathcal{N}_k^j} \bar{m}_{n \to k}^{(l-1)}(x_k) \Bigg ],
\end{align}
which shows that the messages, received at node $k$ and combined with the likelihoods $\ln \phi_k(x_k)$ and $\ln \psi_{kj}(x_k,x_j)$, pass through the following transformation to form the message sent to node $j$
 \begin{equation} \label{eq:g_frak}
f(\boldsymbol{x}) \triangleq \ln \sum_{k} e^{x_k}.
\end{equation}
%
\begin{figure}[]
	\centering 
  \includegraphics[scale=0.55]{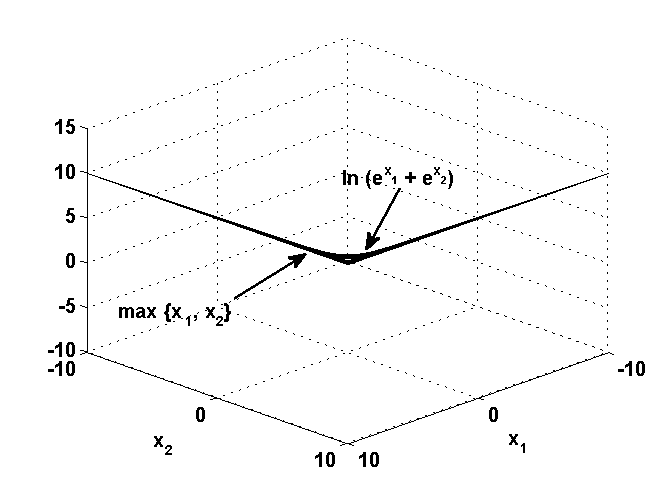} 
  \caption{ $\max \{x_1, x_2\}$ provides a piece-wise linear approximation for $ \ln \left (e^{x_1} + e^{x_2} \right ) $. }
  \label{fig1}
\end{figure}
%
As shown in Fig. \ref{fig1}, due to the highly selective nature of the exponential function, $f(\cdot)$ behaves like a max operator, i.e., 
 \begin{equation} \label{EQ16}
f(\boldsymbol{x}) \approx \max_{k} x_k.
\end{equation}
Consequently, we can approximate the message-update rule in the sum-product algorithm as 
 \begin{align} \label{EQ16}
&\bar{m}_{k \to j}^{(l)}(x_j) \nonumber 
\\
&\approx \max_{x_k} \Bigg  [ \ln \phi_k(x_k) + \ln \psi_{kj}(x_k,x_j) 
+ \sum_{n \in \mathcal{N}_k^j} \bar{m}_{n \to k}^{(l-1)}(x_k) \Bigg ],
\end{align}
which clearly shows that \emph{the message-update rule in the sum-product algorithm is almost the  same as its counterpart in the max-product algorithm. }

Therefore, we expect the max-product algorithm to work as a linear fusion as well. More specifically, we expect to have $\lambda_j = \sum_{k \in \mathcal{V}} w_{jk}\gamma_k$ where $w_{jk}$ denotes the weight of $\gamma_k$ in this linear fusion. In the following section, we formally establish that the max-product algorithm is a distributed linear fusion scheme. 

\section{Analysis of the Max-Product Operation} \label{sec:Analysis}
Performance of a binary hypothesis test is commonly measured by two parameters; the probability of detection and the probability of false alarm. These performance metrics are calculated based on the statistical behavior of the decision variable  $\lambda_j$. Specifically, at node $j$ we have 
\begin{align}
P_{\textup{f}}^{(j)} &= \textup{Pr}\{ \lambda_j > \tau_j \vert x_j = -1\} \label{eq:p_f}, \\
P_{\textup{d}}^{(j)} &= \textup{Pr}\{ \lambda_j > \tau_j \vert x_j = +1\}  \label{eq:p_d},
\end{align}
where $P_{\textup{f}}^{(j)}$ denotes the false-alarm probability of node $j$  and $P_{\textup{d}}^{(j)}$ denotes the corresponding detection probability. 

  Hence, we need to find the probability distribution of $\lambda_j$ to measure the system performance analytically. To realize this goal, we calculate the outcome of each iteration and show that even though the iteration process involves nonlinear transformations, its outcome is a linear combination of the local LLRs. The analysis of the max-product process provided in this section does not require the node variables to be binary-valued. We only use binary-valued $x_i$'s when evaluating the result of the proposed analysis.

 Recall that the local observation at node $k$ is represented by $\phi_k(x_k)$ while the correlation between the observations at nodes $k$ and $j$ is captured by $\psi_{kj}(x_k,x_j)$. Since in the beginning there are no messages received, i.e., $m_{k \rightarrow j}^{(0)}(x_j) = 0$, each node builds its message only based on its own local observation and the correlation of its random variable with the ones of the neighboring nodes. That is, at $l=1$ the messages are created based on
\begin{align}\label{eq:m_kj(1)} 
 m_{k \to j}^{(1)}(x_j) &= \max_{x_k}  \Big [ \ln \phi_k(x_k) + \ln \psi_{kj}(x_k,x_j) \Big] \nonumber \\
 &= \ln \phi_k(\hat{x}_k^{(1)}(x_j)) + \ln \psi_{kj}(\hat{x}_k^{(1)}(x_j),x_j),
\end{align}
where $\hat{x}_k^{(1)}(x_j)$ is found by solving $\frac{\partial}{\partial x_{k}} [ \ln \phi_k(x_k) + \ln \psi_{kj}(x_k,x_j)] = 0$ that leads to 
\begin{equation} \label{eq:x_k(1)} 
\hat{x}_k^{(1)}(x_j) = u_{kj}^{(1)} + v_{kj}^{(1)}x_j,
\end{equation}
where, by using $\frac{\partial}{\partial x_{k}} \ln \phi_k(x_k) = \boldsymbol{s}_k^T (\boldsymbol{y}_k - \xi_k\boldsymbol{s}_k)$, we have 
\begin{align}
u_{kj}^{(1)} &= \frac{2\gamma_k}{E_k} -1 \label{eq:u(1)},\\
v_{kj}^{(1)} &= \frac{2J_{kj}}{E_k} \label{eq:v(1)}.
\end{align}
 Consequently, at the beginning of the iteration, the message sent form node $k$ to node $j$ is a linear function of two components: \emph{i}) the local LLR at node $k$, denoted $\gamma_k$, and \emph{ii}) the realization of the random variable concerned at node $j$, i.e., $x_j$. We see $\hat{x}_k^{(1)}(x_j)$ as the outcome of a maximum-likelihood estimation (MLE) process at node $k$. This estimation provides a point at which the likelihood functions at node $k$ are evaluated to build a message sent to node $j$. Please make sure to distinguish between the MLE performed locally at each node and the MAP estimation discussed earlier. 

As we show in the following, the linear behavior observed in \eqref{eq:x_k(1)}  propagates throughout the entire iteration. Specifically, at the $l$'th iteration, the MLE results at node $k$, which build the messages sent to node $j$, are in the form of linear combinations of $x_j$ and local LLRs obtained at nodes located within less than  $l$ hops from node $j$. Consequently, given $x_j$, the decision variable at node $j$ (i.e., $\lambda_j$) is built by a linear fusion of the local LLRs obtained at node $j$ and at all the nodes located within less than $l$ hops from node $j$. In other words, the hypothesis test result obtained by the max-product algorithm is equivalent to the one obtained by a distributed linear data-fusion scheme whose scope is increased by every iteration. 

We now clarify this observation by solving the iterative optimizations in \eqref{eq:m_kj}. For $l = 2$, we have 
\begin{align}
m_{k \rightarrow j}^{(2)}(x_j)& = \max_{x_k}  \Bigg \{ \ln \phi_k(x_k) + \ln \psi_{kj}(x_k,x_j) \nonumber \\
&+ \sum_{n \in \mathcal{N}_k^j}  \max_{x_n} \bigg [\ln \phi_n(x_n) +  \ln \psi_{nk}(x_n,x_k) \bigg] \Bigg \}, \label{EQ24}
\end{align}
which leads to  
\begin{align} \label{EQ25} 
 m_{k \rightarrow j}^{(2)}(x_j) &= \ln \phi_k(\hat{x}_k^{(2)}(x_j)) +  \ln \psi_{kj}(\hat{x}_k^{(2)}(x_j),x_j) \nonumber \\
 &+ \sum_{n \in \mathcal{N}_k^j}  \Big  [\ln \phi_n(\hat{x}_n^{(1)}(\hat{x}_k^{(2)}(x_j)))  \nonumber \\
 &+ \ln \psi_{nk}(\hat{x}_n^{(1)}(\hat{x}_k^{(2)}(x_j)),\hat{x}_k^{(2)}(x_j)) \Big ],
\end{align}
where $\hat{x}_k^{(2)}(x_j)$ is found by solving 
\begin{align} \label{EQ26} 
\frac{\partial}{\partial x_k} &\Bigg \{ \ln \phi_k(x_k) +  \ln \psi_{kj}(x_k,x_j) \nonumber  \\
&+ \sum_{n \in \mathcal{N}_k^j}  \Big [\ln \phi_n(\hat{x}_n^{(1)}(x_k)) + \ln \psi_{nk}(\hat{x}_n^{(1)}(x_k),x_k) \Big ] \Bigg \} = 0,
\end{align}
which leads to 
\begin{align} \label{eq:x(2)} 
\hat{x}_k^{(2)}(x_j) = u_{kj}^{(2)} + v_{kj}^{(2)}x_j,
\end{align}
where 
\begin{align}
u_{kj}^{(2)} &= \frac{u_{kj}^{(1)} + \frac{1}{E_k}\sum_{n\in \mathcal{N}_k^j} E_n u_{nk}^{(1)}v_{nk}^{(1)}}{1 -\frac{1}{E_k}\sum_{n\in \mathcal{N}_k^j} E_n\left [v_{nk}^{(1)}  \right ]^2 } \label{eq:u(2)},   \\
v_{kj}^{(2)} &= \frac{v_{kj}^{(1)} }{1 -\frac{1}{E_k}\sum_{n\in \mathcal{N}_k^j} E_n\left [v_{nk}^{(1)}  \right ]^2 }\label{eq:v(2)}.
\end{align}

Consequently, $\hat{x}_k^{(2)}(x_j)$ is built as a linear function of $x_j$ plus a linear combination of the local LLRs obtained at node $k$ and at its one-hop neighbors. More specifically, from \eqref{eq:x(2)}, \eqref{eq:u(2)}, and \eqref{eq:v(2)} we see that $\hat{x}_k^{(2)}(x_j)$ is formed as a linear combination of $\gamma_k$ with $\gamma_n$'s for $n \in \mathcal{N}_k^j$. In addition, note that $v_{kj}^{(2)}$ is a constant whereas $u_{kj}^{(2)}$ is a random variable that captures the statistical behavior of the local observations. 

Through iterative calculations for $l=3, 4, ...$, we see that the MLE result $\hat{x}_k^{(l)}(x_j)$ has similar components, i.e., a linear function of $x_j$ plus a linear combination of the local LLRs obtained within less than $l$ hops from node $j$, i.e.,  
\begin{eqnarray} \label{eq:x(k)}
\hat{x}_k^{(l)}(x_j) = u_{kj}^{(l)} + v_{kj}^{(l)}x_j,
\end{eqnarray}
where $v_{kj}^{(l)}$ is a constant and $u_{kj}^{(l)}$ can be expressed as a linear combination of the local likelihood values, i.e., 
\begin{equation}  \label{eq:u(k)}
u_{kj}^{(l)} = \sum_{i \in \mathcal{V}} \omega^{(l)}_{kj}(i) \gamma_i,
\end{equation}
where $\omega^{(l)}_{kj}(i)$ denotes the weight of $\gamma_i$ in this linear combination. Moreover, $\omega^{(l)}_{kj}(i)$ is zero if node $i$ is located more than $l-1$ hops away from node $j$. Therefore, by increasing $l$, we expand the maximum radius around node $j$ within which the local likelihoods are combined to build $u_{kj}^{(l)}$. Consequently, \emph{to include all the local LLRs in the fusion process, the maximum number of iterations does not need to be greater than the length of the longest path in the network graph.} This justifies the observation in \cite{Penna12} where the desired detection performance is achieved by only a few iterations.

 It is worth noting that, one does not need to perform many iterative calculations to see the linearity of the final result. Starting from $m_{k \rightarrow j}^{(1)}(x_j)$ in \eqref{eq:m_kj(1)}, we see that the term $ \ln \phi_k(x_k) + \ln \psi_{kj}(x_k,x_j)$ is concave quadratic in $x_k$. Hence, its partial derivative leads to a linear equation which, in turn, leads to a linear expression for $\hat{x}_k^{(1)}(x_j)$ in terms of $\gamma_k$ and $x_j$. Moreover, in order to build $m_{k \rightarrow j}^{(l)}$ from $m_{k \rightarrow j}^{(l-1)}$ one needs to add some terms, inside the max operator in \eqref{eq:m_kj}, with similar concave quadratic attributes. The only difference is that, these new terms have as their arguments some linear expressions with positive coefficients. Note that $x_k$ in \eqref{eq:m_kj} is calculated by \eqref{eq:x(k)}. Since these linear transformations preserve the concave quadratic nature of the whole expression inside the max operator, the maximum in \eqref{eq:m_kj} is found by solving a linear equation that leads to a linear expression in terms of $u_{kj}^{(l)} $ and $x_j$. 

Based on this observation, we propose a set of formulas to recursively calculate $\hat{x}_k^{(l)}(x_j)$. This calculation is realized by using a quadratic form to represent the messages, which can be expressed as 
\begin{equation}
m_{k \rightarrow j}^{(l)}(x_j) = a_{kj}^{(l)}x_j^2 + b_{kj}^{(l)}x_j.
\end{equation}
The partial derivative of the messages is then a linear expression as
\begin{equation}
\frac{\partial }{\partial x_j} m_{k \rightarrow j}^{(l)}(x_j) = 2a_{kj}^{(l)}x_j + b_{kj}^{(l)}.
\end{equation}
Now, by solving the following equation 
\begin{equation}
\frac{\partial }{\partial x_k} \Big [\ln \phi_k(x_k) + \ln \psi_{kj}(x_k,x_j)  
+ \sum_{n \in \mathcal{N}_k^j} m_{n \rightarrow k}^{(l-1)}(x_k)  \Big ] = 0,
\end{equation}
we link $a_{kj}^{(l)}$ and $b_{kj}^{(l)}$ to $u_{kj}^{(l)}$ and $v_{kj}^{(l)}$ as
\begin{align}
u_{kj}^{(l)} &= \frac{u_{kj}^{(1)} + {2 \over E_k}\sum_{n\in \mathcal{N}_k^j} b_{nk}^{(l-1)}}{1 - {1 \over E_k}\sum_{n\in \mathcal{N}_k^j} E_na_{nk}^{(l-1)} }, \label{eq:u(l)} \\
v_{kj}^{(l)} &= \frac{v_{kj}^{(1)} }{1 - {1 \over E_k}\sum_{n\in \mathcal{N}_k^j} a_{nk}^{(l-1)} } \label{eq:v(l)}.
\end{align}
 Hence, by using $u_{kj}^{(l-1)}$ and $v_{kj}^{(l-1)}$ we calculate $a_{kj}^{(l-1)}$ and $b_{kj}^{(l-1)}$ which are then used to obtain $u_{kj}^{(l)}$ and $u_{kj}^{(l)}$. This recursive calculation starts  from $u_{kj}^{(1)}$ and $v_{kj}^{(1)}$ in \eqref{eq:u(1)} and \eqref{eq:v(1)}.

The fusion weights in \eqref{eq:u(k)} are determined in terms of $v_{kj}^{(l)}$'s.  
As we saw in \eqref{eq:v(1)} and \eqref{eq:v(2)}, $v_{kj}^{(l)}$'s are determined in terms of $J_{kj}$'s which capture the inter-dependencies of the random variables in the MRF. We will further discuss this point and its implications on the system design later. 

 Eqs. \eqref{eq:u(l)} and \eqref{eq:v(l)} indicate that, \emph{the higher the received SNR level at node $k$, the lower the impact of other nodes on the data sent from node $k$ to node $j$.} Hence, node $k$ relies more on its own local observation when it is operating under good SNR conditions. Otherwise, it relies more on the data received from its neighbors. In addition, according to \eqref{eq:u(2)} and \eqref{eq:v(2)}, each LLR received from a neighbor is scaled by the SNR level perceived at that neighbor. Consequently, the message-update rule in \eqref{eq:m_kj} works like a maximal-ratio combining (MRC) scheme. 

Since the outcomes of the MLEs are derived in closed form, we can now see their impact on the binary hypothesis test. 
To this end, we show that $\delta_{k \to j}^{(l)}$ is a linear combination of the local LLRs. First note that for all $k,n$ we have 
\begin{align} 
\ln \phi_k(u+v) - \ln \phi_k(u-v) = 2v  \left[\gamma_k - \frac{E_k}{2}(u+1)  \right] & \label{eq:phi(u+v)}, \\ 
\ln \psi_{nk}(u_1+v_1,u_2+v_2) - \ln \psi_{nk}(u_1-v_1,u_2-v_2) &= \nonumber \\
2J_{nk}(u_1v_2+u_2v_1) \label{eq:psi(u+v)},
\end{align}
which are linear expressions in $u, u_1$, and $u_2$. 

Then, recall that $\hat{x}_k^{(l)}(x_j = \pm 1) = u_{kj}^{(l)} \pm v_{kj}^{(l)} $. Consequently, $\delta_{k \to j}^{(l)}$ in \eqref{eq:delta_kj} contains expressions, in the form of \eqref{eq:phi(u+v)} and \eqref{eq:psi(u+v)}, that are linear functions of $u_{kj}^{(l)}$'s. 
To clarify this observation, we focus on $l=2$ here. A similar argument can be made for $l>2$. We see that, 
\begin{align}
& \delta_{k \to j}^{(2)} = m_{k \rightarrow j}^{(2)}(x_j= +1) - m_{k \rightarrow j}^{(2)}(x_j= -1) = \nonumber \\
& \ln \phi_k(\hat{x}_k^{(2)}(+1)) - \ln \phi_k(\hat{x}_k^{(2)}(-1)) \nonumber \\
&+ \ln \psi_{kj}(\hat{x}_k^{(2)}(+1),+1) - \ln \psi_{kj}(\hat{x}_k^{(2)}(-1),-1)\nonumber \\
&+ \sum_{n \in \mathcal{N}_k^j} \Big [\ln \phi_n (\hat{x}_n^{(1)}(\hat{x}_k^{(2)}(+1))) - \ln \phi_n(\hat{x}_n^{(1)}(\hat{x}_k^{(2)}(-1))) \Big ]\nonumber \\
&+ \sum_{n \in \mathcal{N}_k^j} \Big [ \ln \psi_{nk}(\hat{x}_n^{(1)}(\hat{x}_k^{(2)}(+1)),\hat{x}_k^{(2)}(+1)) \nonumber \\
&~~~~~~~~~~~-\ln \psi_{nk}(\hat{x}_n^{(1)}(\hat{x}_k^{(2)}(-1)),\hat{x}_k^{(2)}(-1)) \Big], \label{eq:delta(2)}
\end{align}
where 
\begin{align}
\hat{x}_k^{(2)}(\pm 1) &= u_{kj}^{(2)} \pm v_{kj}^{(2)}, \label{EQ37}\\
\hat{x}_n^{(1)}(\hat{x}_k^{(2)}(\pm 1)) &= u_{nk}^{(1)} + v_{nk}^{(1)}\hat{x}_k^{(2)}(\pm 1) \nonumber \\
&= u_{nk}^{(1)} + v_{nk}^{(1)}u_{kj}^{(2)} \pm v_{nk}^{(1)}v_{kj}^{(2)}. \label{EQ38}
\end{align}
Comparing \eqref{eq:delta(2)} to \eqref{eq:phi(u+v)} and \eqref{eq:psi(u+v)} makes it clear that, \eqref{eq:delta(2)} is a linear combination of $u_{nk}^{(1)}$ and  $u_{kj}^{(2)}$, for $k \in \mathcal{N}_j$ and $n \in \mathcal{N}_k^{j}$. Since  $u_{nk}^{(1)}$ and  $u_{kj}^{(2)}$ are, respectively, linear combinations of the local likelihoods $\gamma_k$'s for $k \in \mathcal{N}_j$ and $\gamma_n$'s for $n \in \mathcal{N}_k^{j}$, we conclude that $\lambda_j^{(2)}$ is a linear combination of $\gamma_k$'s for $k \in \{j\} \cup \mathcal{N}_j$ and $\gamma_n$'s for $n \in \mathcal{N}_k^{j}$.

Through similar arguments, we can show that the resulting decision variable after $l$ iterations is constructed as a linear combination of the local likelihoods, i.e., 
\begin{equation} \label{EQ39}
\lambda_j^{(l)} = \sum_{i \in \mathcal{M}_j^{(l)}} w^{(l)}_{ji} \gamma_i,
\end{equation}
where $w^{(l)}_{ji}$'s denote the weights in this linear combination while $\mathcal{M}_j^{(l)}$ denotes the set of indices referring to node $j$ and all its neighbors within its $(l-1)$-hop distance. Consequently, given enough time or when the max-product algorithm converges to a fixed point, the decision variable is built as 
\begin{equation} \label{eq:lambda_j_maxprod}
\lambda_j = \sum_{i \in \mathcal{V}} w_{ji} \gamma_i,
\end{equation}
where $\lambda_j \triangleq \lim_{l \to \infty} \lambda_j^{(l)}$. We can summarize these observations in the following proposition. 

\emph{Proposition I}: The max-LLRs obtained by running the max-product algorithm, over a network described by the factor graph in \eqref{eq:MRF}  where $ \ln \phi_k(x_k) + \ln \psi_{kj}(x_k,x_j)$ is concave quadratic in $x_k$, are built as linear combinations of the local LLRs in that network. Moreover, at the $l$'th iteration, each node combines local LLRs from its neighbors located within less than $l$ hops away from itself. 

We know that $J_{kj}$'s specify the factor graph which models the stochastic behavior of the network. The proposed analysis shows that the fusion weights in \eqref{eq:p(x|y)_2} are determined in terms of $J_{kj}$'s. Therefore, finding the optimal $J_{kj}$'s to best represent the network behavior is equivalent to optimizing the fusion weights in \eqref{eq:lambda_j_maxprod}. This observation gives us a deeper insight into the impact of the MRF parameters on the system performance and enables us to offer our second proposition as follows. 

\emph{Proposition II}: Learning the parameters of the pairwise factor graph in \eqref{eq:p(x|y)_2} to best represent the statistical correlations in a network of distributed agents and running the max-product algorithm based on that graph can be viewed as the optimization of a distributed linear data-fusion scheme in that network. 

Linear data-fusion has been extensively investigated in the literature. For the completeness of presentation, we briefly explain how to realize optimal linear data-fusion in the following section. An interested reader may refer to \cite{Quan08, Quan10, Abdi14} for more comprehensive discussions. We have explained in detail the proposed optimization framework in \cite{Abdi20} where we optimize the sum-product algorithm. In the following section, we discuss why that framework can be applied to the max-product algorithm as well.  

\section{Linear Data-Fusion} \label{sec:linearFusion}
The fact that the decision variable $\lambda_j$ is the result of a linear fusion facilitates the statistical analysis of the system behavior and optimizing its performance. Given the status of $x_i$'s, the local LLRs follow Gaussian distributions which means that the decision variable $\lambda_j$ follows a Gaussian distribution and we only need its first- and second-order statistics to derive its probability distribution. We can find the impact of fusion weights (i.e., $w_{jk}$'s) on the system performance by noting that they determine the contribution of each node on the mean and variance of $\lambda_j$. The system false-alarm and detection probabilities at node $j$ depend not only on the state of $x_j$, but also, in general, on the state of all other $x_i$'s being sensed throughout the entire network. Consequently, based on the total probability theorem, we have  
\begin{align} 
& g_{j}(\tau_j,v) \triangleq \textup{Pr}\{\lambda_j > \tau_j \vert x_j = v\} \nonumber \\
 &= \sum_{\boldsymbol{b} \in \{-1,1\}^{N-1}}\textup{Pr}\{\lambda_j > \tau_j \vert \boldsymbol{x}_{(j)} = \boldsymbol{b}, x_j = v\}p_{\boldsymbol{x}_{(j)} |x_j}(\boldsymbol{b}|v) \nonumber\\
 &= \sum_{\boldsymbol{b} \in \{-1,1\}^{N-1}} Q \left(\frac{\tau_j - \eta_{j,v}(\boldsymbol{b})}{\sigma_{j,v}(\boldsymbol{b})} \right )p_{\boldsymbol{x}_{(j)} |x_j}(\boldsymbol{b}|v), \label{g}
\end{align}
where $\boldsymbol{x}_{(j)} \triangleq [x_1,x_2, ..., x_{j-1}, x_{j+1}, x_{j+2}, ..., x_N]$, $p_{\boldsymbol{x}_{(j)}|x_j}(\boldsymbol{b}|v) \triangleq \textup{Pr}\{\boldsymbol{x}_{(j)} = \boldsymbol{b} \vert x_j = v\}$, and for $v=-1,1$, $\eta_{j,v}(\boldsymbol{b}) \triangleq E[\lambda_j \vert \boldsymbol{x}_{(j)}= \boldsymbol{b}, x_j = v]$ and $\sigma_{j,v}^2(\boldsymbol{b}) \triangleq \textup{Var}[\lambda_j \vert \boldsymbol{x}_{(j)} = \boldsymbol{b}, x_j = v]$. $Q(x) \triangleq \int_{x}^{\infty} \frac{1}{\sqrt{2\pi}} e^{-z^2/2} dz$ is the so-called $Q$-function. 
Note that $\boldsymbol{x}_{(j)} \in \{-1,1\}^{N-1}$ contains all $x_i$'s except for $x_j$. It is clear that $P_\textup{f}^{(j)} = g_j(\tau_j,-1)$ and $P_\textup{d}^{(j)} = g_j(\tau_j,1)$. By solving $g_j(\tau_j,-1) = \alpha$ or $g_j(\tau_j,1) = \beta$ we obtain a value for $\tau_j$ which guarantees the false-alarm or detection probability at node $j$ be, respectively, equal to $\alpha$ or $\beta$. We have discussed how to find the detection threshold to guarantee a predefined performance level in \cite{Abdi20}.

\subsection{Centralized Linear Fusion} \label{sec:OptimalLinearFusion}
In a centralized distributed detection \cite{Akyildiz11}, the local sensing outcomes are constantly reported to a so-called fusion center (FC) which is usually a more powerful node like a base station or an access point. The FC uses the received information from the cooperating nodes to estimate the statistics required for a linear fusion scheme and then directly combines the received local sensing results into a global decision variable. In this fusion process, the vector of decision variables is built as 
\begin{equation} \label{EQ52}
\boldsymbol{\lambda} = \boldsymbol{W}\boldsymbol{\gamma},
\end{equation}
where $\boldsymbol{W}$ denotes the weighting coefficients, see \eqref{eq:lambda_j_maxprod}. 

Now, the problem is to find the optimal fusion weights and detection thresholds to have the best detection performance. A linear fusion scheme can be optimized by assigning rewards to the detection and costs to the false-alarm incidents. In particular, we assume that the system obtains reward $r_i$ for performing a correct detection at node $i$ and incurs cost $c_i$ when a false alarm happens. Therefore, the average reward obtained by the system regarding the detection performance at node $i$ is $r_i P_{\textup{d}}^{(i)}$ whereas the average cost of false alarms happening at that node is $c_i P_{\textup{f}}^{(i)}$. Consequently, the aggregate reward obtained throughout the entire network is stated as $R(\boldsymbol{\lambda}, \boldsymbol{\tau}) \triangleq \mathbf{r}^T \mathbf{P}_{\textup{d}}$ where $\mathbf{r} = [r_1, ..., r_K]^T$ and  $\mathbf{P}_{\textup{d}} = [P_{\textup{d}}^{(1)}, ..., P_{\textup{d}}^{(K)}]^T$ while the aggregate cost of false alarms is taken into account by $C(\boldsymbol{\lambda}, \boldsymbol{\tau}) = \mathbf{c}^T \mathbf{P}_f$ where $\mathbf{c} = [c_1, ..., c_K]^T$ and  $\mathbf{P}_{\textup{f}} = [P_{\textup{f}}^{(1)}, ..., P_{\textup{f}}^{(K)}]^T$. Accordingly, the system performance optimization is formulated as 
\begin{gather*}\label{P1}
 \operatorname*{max}_{\mathbf{W},\boldsymbol{\tau} } R(\boldsymbol{\lambda}, \boldsymbol{\tau}), \tag{P1} \\ 
\begin{matrix}
\textup{s.t.,} & C(\boldsymbol{\lambda}, \boldsymbol{\tau}) \le C_0, & \mathbf{P}_\textup{f}(\boldsymbol{\lambda}, \boldsymbol{\tau}) \le \boldsymbol{\alpha}, & \mathbf{P}_\textup{d}(\boldsymbol{\lambda}, \boldsymbol{\tau}) \ge \boldsymbol{\beta},
\end{matrix}
\end{gather*}
where $C_0$ denotes the maximum cost allowed while $\boldsymbol{\alpha} = [\alpha_1, ..., \alpha_K]^T$ and  $\boldsymbol{\beta} = [\beta_1, ..., \beta_K]^T$ denote the per-node constraints the system has to meet when performing the hypothesis test. That is, we optimize the aggregate system performance while maintaining the constraints $P_{\textup{f}}^{(j)} \le \alpha_j$ and $P_{\textup{d}}^{(j)} \ge \beta_j$ for $j = 1, ..., N$. The optimization problem in \eqref{P1} is solved in \cite{Quan09} in the context of distributed multiband spectrum sensing in CR networks. 
 
\subsection{Decentralized Linear Fusion} \label{sec:DecDistInf}
The message-passing algorithms are of special interest in decentralized distributed settings where there is no FC and the network has to conduct the detection process based on limited computation and communication resources offered only by the sensing nodes. Here we discuss how to optimize the message-passing process in such a design scenario. 

Since the max-product algorithm works like the sum-product algorithm, we first explain how to derive the optimal detection performance by optimizing the sum-product algorithm. This optimization framework is based on the fact that the resulting linear fusion favors the LLRs received from shorter distances, especially, the ones generated at the one-hop neighbors. Then, we show that the linear fusion realized by the max-product algorithm has the same property. Therefore, the same optimization framework gives the optimal detection performance of the max-product algorithm as well. 

 Eq. \eqref{eq:lambda_j_linear}  reveals the effect of the network topology
on how the fusion coefficients are arranged by a sum-product iteration. Specifically,  for the one-hop neighbors of node $j$ we have one coefficient
$c_{jk}$ affecting the local LLRs received, for the two-hop neighbors we have two coefficients $c_{jk}c_{kn}$ and so on. Since $|c_{jk}|<1$, the system favors LLRs received through the shortest paths when building the decision variables. Moreover, the impact of $\gamma_n$ on $\lambda_j$, which depends on the correlation between $x_n$ and $x_j$, is determined by multiplying two factors $c_{kn}$ and $c_{jk}$ corresponding, respectively, to the link from node $n$ to node $k$ and the link from node $k$ to node $j$. Accordingly, we decompose the problem of optimizing $c_{jk}$'s into $N$ small optimizations, each carried out locally at a sensing node, which collectively lead to a near-optimal performance in a decentralized distributed network configuration. 

In the proposed optimization framework each node is focused on the fusion of the LLRs received from its one-hop neighbors by using the following approximation \cite{Abdi20}
\begin{equation}\label{lambda_j_neighbors}
\bar{\lambda}_j  \approx \gamma_j + \sum_{k \in \mathcal{N}_j}c_{jk}\gamma_k,
\end{equation}
which is used to formulate the local optimizations as 
\begin{gather*}\label{P2}
 \operatorname*{max}_{\boldsymbol{c}_j, \tau_j } P_{\textup{d}}^{(j)}(\bar{\lambda}_j), \tag{P2} \\ 
\begin{matrix}
\textup{s.t.,} & P_{\textup{f}}^{(j)}(\bar{\lambda}_j) \le \alpha_j,\\
& |c_{jk}| < \frac{1}{\operatorname*{max}_{n} |\mathcal{N}_n|-1}, \forall k \in \mathcal{N}_j,
\end{matrix}
\end{gather*}
where we maximize the detection probability at node $j$ while maintaining its false-alarm probability below the predefined threshold $\alpha$. This optimization  is based on the Neyman-Pearson method \cite{Kay93}. $P_{\textup{d}}^{(j)}$ and $P_{\textup{f}}^{(j)}$ are derived approximately by 
\begin{align}\label{gApprox}
g_j(\tau_j,v) \approx \sum_{\boldsymbol{b} \in \{0,1\}^{\left|\mathcal{N}_j\right|}} Q \left(\frac{\tau_j - \tilde{\eta}_{j,v}(\boldsymbol{b})}{\tilde{\sigma}_{j,v}(\boldsymbol{b})} \right )p_{\tilde{\boldsymbol{x}}_{(j)} |x_j}(\boldsymbol{b}|v),
\end{align}
where $\tilde{\boldsymbol{x}}_{(j)}$ is a vector which contains $x_i$'s with $i \in \mathcal{N}_j$ while for $v = 0,1$, we have $\tilde{\eta}_{j,v}(\boldsymbol{b}) \triangleq E[\bar{\lambda}_j \vert \tilde{\boldsymbol{x}}_{(j)} = \boldsymbol{b}, x_j = v]$ and $\tilde{\sigma}_{j,v}^2(\boldsymbol{b}) \triangleq \textup{Var}[\bar{\lambda}_j \vert \tilde{\boldsymbol{x}}_{(j)} = \boldsymbol{b}, x_j = v]$. Note that, $\left|\mathcal{N}_j\right|$ denotes the number of one-hop neighbors of node $j$ while $\tilde{\eta}_{j,v}$ and $\tilde{\sigma}_{j,v}$ denote an estimation of the first- and second-order conditional statistics of $\bar{\lambda}_j$ given the value of $x_j$ and its immediate neighbors. This optimization can be solved by the blind adaptation method provided in \cite{Abdi20} when the required statistics are not available a priori. The last constraint in \eqref{P2} is imposed by the \emph{contracting mapping principle} \cite{Mooij07} which guarantees the convergence of the message-passing iteration. 

By solving \eqref{P2}, the optimal $c_{jk}$'s are found in terms of the correlations between the LLRs made at nodes $j$ and $k$. Consequently, if node $n$ is connected to node $j$ through node $k$, the correlation between $x_j$ and $x_k$ is captured in $c_{jk}$ while the correlation between $x_k$ and $x_n$ is accounted for by $c_{kn}$. Hence, both of the correlations concerned are taken into account in the system design by the multiplication $c_{jk}c_{kn}$ in \eqref{eq:lambda_j_linear} while each node sees its immediate neighbors only when optimizing its own fusion coefficients. This fusion process is inline with the Markovian structure of the factor graph. Recall that, the correlation between $x_n$ and $x_j$ is accounted for in \eqref{eq:MRF} by two factors $\psi_{k,n}(x_k,x_n)$ and $\psi_{jk}(x_j,x_k)$ multiplied together within $p(\boldsymbol{x} \vert \boldsymbol{Y})$.

The same approach can be used to optimize the max-product algorithm. The reason is that the local LLRs received from the one-hop neighbors have dominant effects on the decision variables build by the max-product operation. This is a behavior we saw earlier in the sum-product algorithm. To clarify this  observation, when formulating the decision variable $\lambda_j^{(l)}$, we use the fact that  $v_{kj}^{(1)}$ is proportional to $\frac{1}{K}$ and, assuming the number of samples $K$ to be large \cite{Quan08, Quan10, Abdi14}, we can see that the system favors data received from closer distances. Again, we focus on $l=2$ for simplicity. By using \eqref{eq:phi(u+v)} and \eqref{eq:psi(u+v)} while approximating the terms proportional to $1 \over K$ by zero, we have  
\begin{align}
&\ln \phi_k(\hat{x}_k^{(2)}(+1)) - \ln \phi_k(\hat{x}_k^{(2)}(-1)) \nonumber \\
&\approx v_{kj}^{(1)} \left[ \gamma_k - \frac{E_k}{2} \left (u_{kj}^{(1)} +1 \right )  \right] = 0, \label{EQ61}\\
&\ln \psi_{kj}(\hat{x}_k^{(2)}(+1),+1) - \ln \psi_{kj}(\hat{x}_k^{(2)}(-1),-1) \approx 2J_{kj} u_{kj}^{(1)}, \label{EQ62}\\
&\ln \phi_n (\hat{x}_n^{(1)}(\hat{x}_k^{(2)}(+1))) - \ln \phi_n(\hat{x}_n^{(1)}(\hat{x}_k^{(2)}(-1))) \nonumber \\
&= v_{nk}^{(1)}v_{kj}^{(2)} \left[\gamma_n - \frac{E_n}{2} \left (u_{nk}^{(1)} + v_{nk}^{(1)}u_{kj}^{(2)}+1 \right) \right] \approx 0, \label{EQ63}\\
&\ln \psi_{nk}(\hat{x}_n^{(1)}(\hat{x}_k^{(2)}(+1)),\hat{x}_k^{(2)}(+1)) \nonumber \\
& -\ln \psi_{nk}(\hat{x}_n^{(1)}(\hat{x}_k^{(2)}(-1)),\hat{x}_k^{(2)}(-1)) \nonumber \\
& = 2J_{kj}\left [(u_{nk}^{(1)} + v_{nk}^{(1)}u_{kj}^{(2)})v_{kj}^{(2)} + u_{kj}^{(2)}v_{nk}^{(1)}  \right ] \approx 0. \label{EQ64}
\end{align}
By plugging these approximations into \eqref{eq:delta(2)}, we derive the fusion result in \eqref{eq:lambda_j_linear} as\footnote{We have merged 2 into $J_{kj}$.}
\begin{equation}\label{eq:lambdaApprox}
\lambda_j \approx \gamma_j + \sum_{k \in \mathcal{N}_j} J_{kj}\gamma_k, 
\end{equation}
which shows that firstly, the decision variable of each node can be derived approximately by linearly combining its local LLR with the LLRs obtained at its one-hop neighbors. Secondly, the likelihoods are scaled in this combination by $J_{kj}$'s which capture the correlation between the node variables. 
And, thirdly, we only need the statistics of the one-hop neighbors here to analyze the stochastic behavior of $\lambda_j$. 

Since $J_{kj}$'s are our degrees of freedom in this design, \eqref{lambda_j_neighbors} and \eqref{eq:lambdaApprox} are the same from an optimization point of view. Note that, $c_{jk}$ is a monotonically-increasing function of $J_{kj}$. 

To sum up, the max-product and sum-product operations provide almost the same message-passing algorithms in which the local LLRs are combined linearly to build the decision variables while the closer neighbors having a more significant contribution on the decision variables obtained. Consequently, the optimal detection performance of a max-product-based system can be achieved by the optimal linear message-passing algorithm defined by \eqref{eq:delta_sumprod} in which the coefficients are obtained by \eqref{P2}.

 \subsection{Discussion} \label{subsec:Discussion}
 In case there is a node at which the maximum of $q(x_j \vert \boldsymbol{y})$ is not attained at a unique value, then the hypothesis test conducted by using the max-marginals is not necessarily equal to the MAP estimation in \eqref{mapEq}. This means that compared to the MAP estimation we may have some extra error in the system performance. However, by using the proposed analysis, now we can control this error and even minimize it. In addition, this performance optimization can be realized with low computational complexity and in a distributed setting. Hence, in practice, the proposed message-passing process  provides performance guarantees and ease of implementation not typically available when dealing with a generic MAP estimation process. In the following section, we show that the proposed detector closely achieves the optimal performance level. 
 
Moreover, the detection performance is affected by the parameters adopted for the MRF and finding the optimal or even near-optimal values for those parameters is certainly a challenge, see e.g., \cite{Lokhov18}. Based on the proposed analysis, now we can optimize the resulting data-fusion process and that is equivalent to optimizing the parameters of the MRF. This was not possible before. 

\begin{figure}[]
	\centering 
  \includegraphics[scale=0.40]{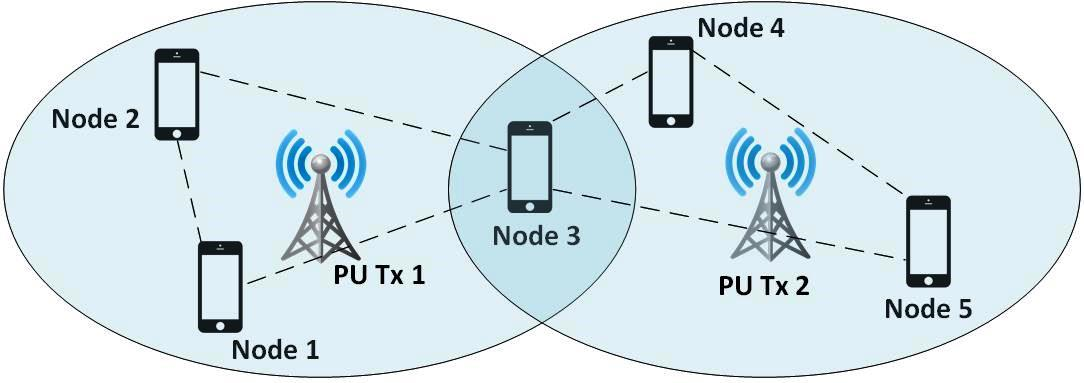} 
  \caption{The network configuration considered in the simulations. Five sensing nodes cooperate to find transmission opportunities within the spectrum bands allocated to two primary transmitters. The dashed lines depict the links between the sensing nodes through which the distributed detection is conducted.    }
  \label{fig:network}
\end{figure}

\section{Numerical Results}\label{sec:Simulations}
In this section, we verify our analysis by computer simulations. Fig. \ref{fig:network} demonstrates the network configuration considered in our simulations. We use the same network structure as in \cite{Abdi20}. Specifically, five SUs are cooperating in an ad-hoc setting via a parallel message-passing iteration, in sensing the radio spectrum to find vacant bands temporarily not in use by two PU transmitters. Nodes 1 and 2 are located within the range of PU transmitter 1 and nodes 4 and 5 are located within the range of PU transmitter 2 while node 3 can receive signals from both of the PU transmitters. The distributed detection is conducted through the one-hop links depicted in Fig. \ref{fig:network} by dashed lines. We realize a spatially-correlated occupancy pattern by making the PU transmitters exhibit correlated random on and off periods. This is an extension of the simulation scenario in \cite{Penna12} where one of the PU transmitters is constantly on while the other one is off all the time.

The spatial diversity in the network structure is accounted for by assigning different SNR levels to different nodes. Specifically, the SNR levels, in dB, of the signals received from PU transmitter 1 at nodes 1, 2, and 3 are $\rho+\Delta \rho$, $\rho$, and $\rho-\Delta \rho$, while the SNR levels of the signals received from PU transmitter 2 at nodes 3, 4, and 5 are $\rho$, $\rho - \Delta \rho$, and $\rho+\Delta \rho$, respectively. Consequently, $\rho$ denotes the average SNR level in the network while $\Delta \rho$ measures the level of SNR dispersion among the network nodes. 

\begin{figure}[]
	\centering 	
	\begin{subfigure}{.45\textwidth}
	  \centering 
  \includegraphics[scale=0.34]{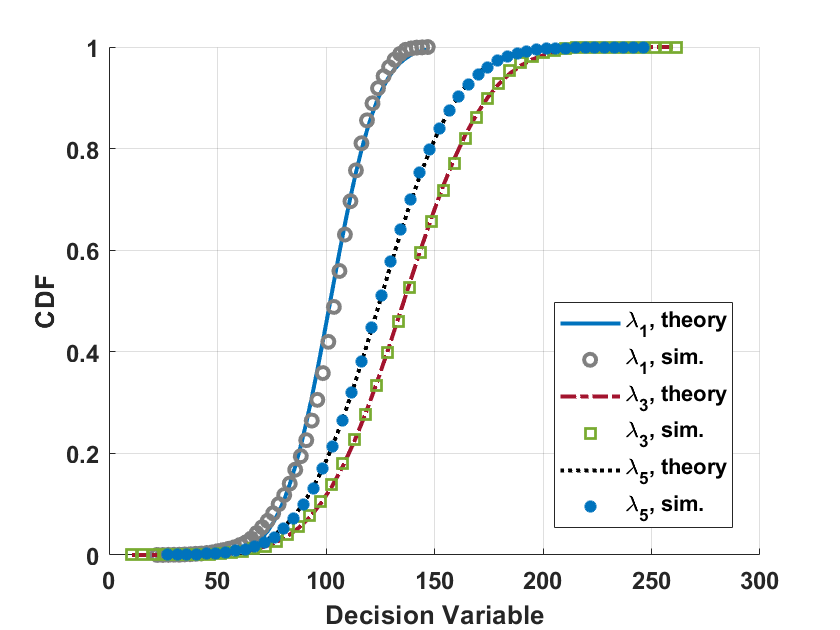} 
  \caption{PU Tx 1 is active and PU Tx 2 is inactive.}
  \label{fig:cdfsCoherent10} 
  \end{subfigure}  
  \begin{subfigure}{.45\textwidth}
    \centering 
  \includegraphics[scale=0.34]{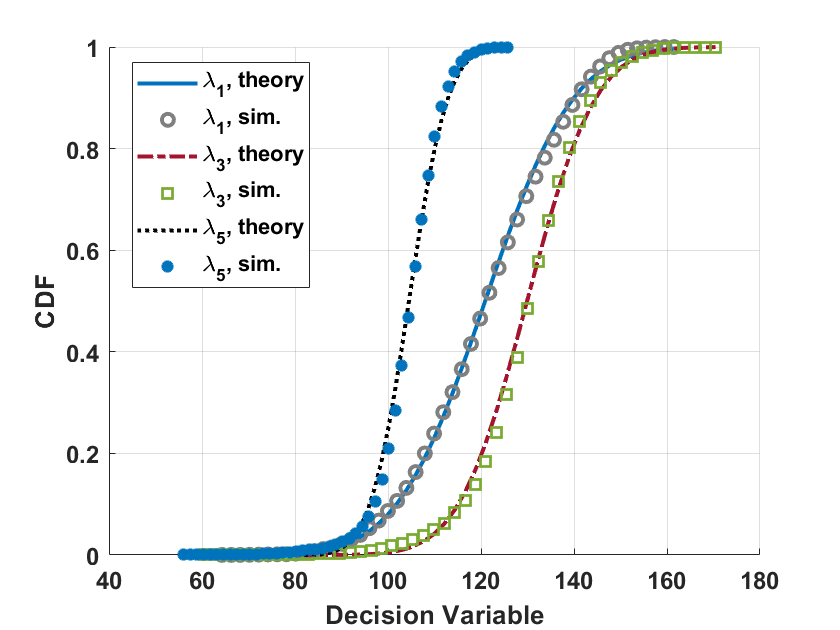} 
  \caption{Both PU transmitters are active.}
  \label{fig:cdfsCoherent11} 
  \end{subfigure}  
  \caption{Cumulative distribution functions of the decision variables built by the max-product algorithm while coherent detection is used as the local sensing method. The resulting decision variables follow Gaussian distributions given the status of the PU transmitters. }
\end{figure}
\begin{figure}[]
	\centering 	
	\begin{subfigure}{.45\textwidth}
	  \centering 
  \includegraphics[scale=0.34]{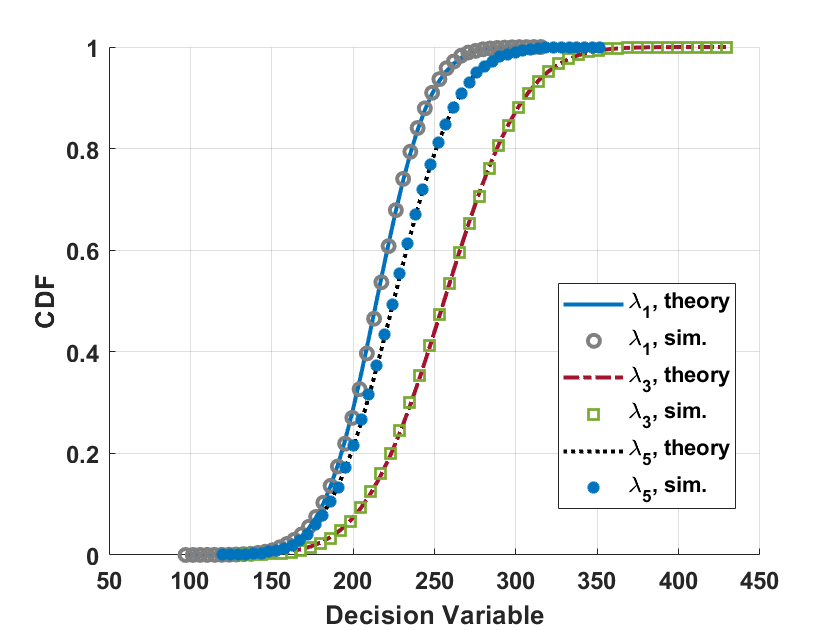} 
  \caption{PU Tx 1 is active and PU Tx 2 is inactive.}
  \label{fig:cdfsED10} 
  \end{subfigure}  
  \begin{subfigure}{.45\textwidth}
    \centering 
  \includegraphics[scale=0.34]{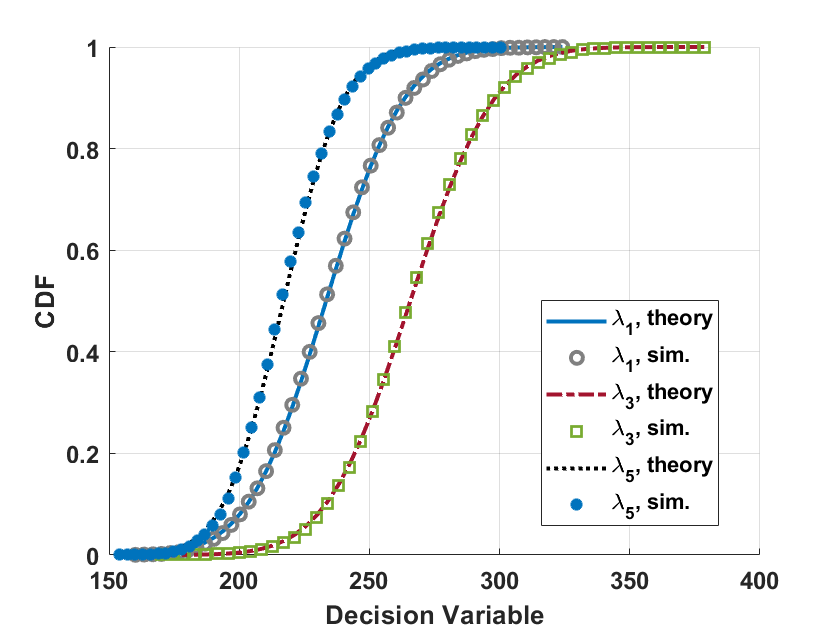} 
  \caption{Both PU transmitters are active.}
  \label{fig:cdfsED11} 
  \end{subfigure}  
  \caption{Cumulative distribution functions of the decision variables built by the max-product algorithm while energy detection is used as the local sensing method. The resulting decision variables follow Gaussian distributions given the status of the PU transmitters. }
\end{figure}

We first study the statistical behavior of the decision variables built by the max-product algorithm. Specifically, we show that, given the status of the PU transmitters, $\lambda_i$'s follow Gaussian distributions when coherent detection or energy detection is used in the sensing nodes. Figs. \ref{fig:cdfsCoherent10} and \ref{fig:cdfsCoherent11} depict the cumulative distribution functions (CDF) of the decision variables obtained at nodes 1, 3, and 5 when coherent detection is used at each node by processing 100 samples of the received signal, i.e., when $K=100$. In this simulation, $\rho = -5$ dB and $\Delta \rho = 1$ dB while $J_{kj}$ values are randomly drawn, with a uniform distribution, from $(0,100)$. For a given realization of $J_{kj}$'s, each data point is obtained by averaging over 10,000 detection outcomes. Fig. \ref{fig:cdfsCoherent10} depicts the case in which only PU transmitter 1 is active while in the case depicted in Fig. \ref{fig:cdfsCoherent11} both PU transmitters are active. For each decision variable, we have provided two curves. The curves labeled ''sim.'' depict the CDFs of the decision variables observed in our simulations whereas the ones labeled ''theory'' depict Gaussian CDFs fitted to the behavior of those decision variables. We can now clearly see the Gaussian behavior in the decision variables and this behavior validates our argument regarding the linearity of the max-product operation. 

Such a linear behavior is also demonstrated in Figs. \ref{fig:cdfsED10} and \ref{fig:cdfsED11} where energy detection is used as the local sensing method. Again, we see that the decision variables obtained by the max-product algorithm follow Gaussian distributions. The reason is that, when we replace matched filtering by energy detection, we do not alter the structure of the max-product algorithm which leads to a linear combination of the local sensing outcomes. We only change the local sensing outcomes exchanged by the sensing nodes. Since the local sensing outcomes produced by energy detection are Gaussian random variables, a linear combination of those outcomes follows a Gaussian distribution as well. 

\begin{figure}[]
	\centering 	
	  \centering 
  \includegraphics[scale=0.40]{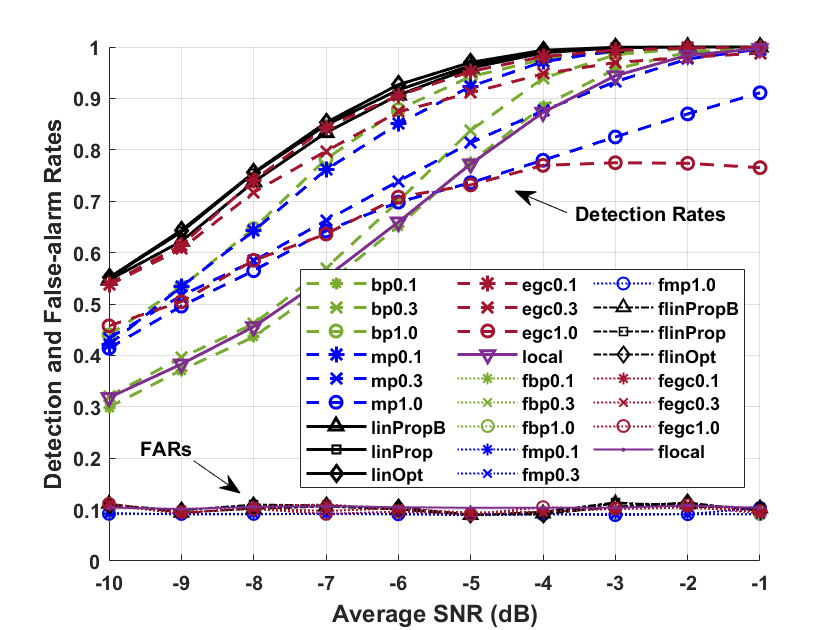} 
  \caption{Performance of different detection methods under different average SNR levels. The SNR dispersion level is $\Delta \rho = 0.1 \rho$. The detection and false-alarm rates are obtained by averaging the corresponding rates over all five sensing nodes in the network. }
  \label{fig:PdPf1} 
  \end{figure}

Since we have now established that the max-product algorithm is a linear fusion method, we know that its performance level is bounded from above by the performance of the optimal linear fusion scheme. We also know from \cite{Quan08} that the performance of the optimal linear fusion is very close to that of the optimal detector, i.e., likelihood-ratio test. In addition, we expect that the linear message-passing framework in \eqref{P2}, which closely achieves the optimal detection performance, outperforms both the max-product and sum-product algorithms. We also expect the proposed method to outperform the equal-gain combining (EGC) method, which is a linear fusion scheme that treats all local sensing outcomes equally \cite{Ma09}. All these expectations are confirmed by the results depicted in Fig. \ref{fig:PdPf1}.

In Fig. \ref{fig:PdPf1}, we compare the performance levels of all these methods under different SNR levels. Specifically, for the SNR dispersion level of $\Delta \rho = 0.1 \rho$, the average detection rate of the different methods discussed are depicted in Fig. \ref{fig:PdPf1} vs. different values of the average SNR level $\rho$ while their false-alarm rates are fixed at 0.1. The curves in Fig. \ref{fig:PdPf1} measure the average of the detection and false-alarm rates over all of the five sensing nodes in the network while each data point is obtained by averaging over 20,000 sensing outcomes. Each sensing outcome is calculated by processing 100 received signal samples at each node for energy detection. A window of $T = 2500$ time slots is used for training the sum-product, max-product, and the proposed linear message-passing algorithms. We realize EGC by $c_{jk} = c_0$, for all $(j,k) \in \mathcal E$, where $c_0$ is a constant. 

In Fig. \ref{fig:PdPf1}, we depict the average detection rate achieved by the max-product algorithm for $\zeta = 0.1, 0.3, 1.0$ in \eqref{lrnFact} by curves labeled, respectively, as ''mp0.1'', ''mp0.3'' and ''mp1.0'' while for the corresponding curves regarding the sum-product algorithm we use labels ''bp0.1'', ''bp0.3'' and ''bp1.0''. The optimal linear fusion is labeled ''linOpt'' and the proposed linear message-passing algorithm in \cite{Abdi20} is labeled ''linProp'' while the detection rate of its blind version is labeled ''linPropB''. EGC is realized in our simulations for $c_0 = 0.1, 0.3, 1.0$ whose detection rates in Fig. \ref{fig:PdPf1} are labeled, respectively, by ''egc0.1'', ''egc0.3'' and ''egc1.0''. The label ''local'' refers to the local sensing method, which is energy detection here, performed individually at each node and without cooperating with other nodes. To refer to the false-alarm rate (FAR) of a specific method we put an ''f'' in the beginning of the label already used for the detection rate of that method. For instance, we use ''fmp0.1'' to refer to the FAR of the detection method labeled ''mp0.1''. 

 The blind optimization in \cite{Abdi20} is conducted based on an offline iterative estimation of the required statistics where each node, say node $j$, processes data received from its neighbors to derive a reliable $\hat x_j$, which is then used in estimating the required statistics, i.e., $\textup E[\gamma_k | x_j]$ and $\textup{cov}(\gamma_i,\gamma_k | x_j)$ for $i,k \in \mathcal N_j$. This estimation can be improved by adding a majority rule \cite{Chaudhari12} to the iteration. Specifically, at each iteration node $j$ corrects its decision by applying the majority rule on $\hat x_j$ and the estimates of $x_j$ inferred from $\gamma_k$'s generated by its neighbors. Those estimates are obtained by thresholding $\gamma_k$'s received at node $j$ while the thresholds are simply updated by using the mean and variance of the same received data. Iterative correction of $\hat x_j$'s significantly increases the effectiveness of our offline adaptation and enables us to achieve near-optimal results in very low SNR regimes as shown in Fig. \ref{fig:PdPf1}. 
 
As we expected, Fig. \ref{fig:PdPf1} shows that the optimal linear fusion scheme exhibits the highest detection rates in all the SNR levels considered. Note however that, this method requires the first- and second-order statistics of all the local sensing outcomes to be available a priori \cite{Quan08}. The proposed linear message-passing algorithm closely achieves the optimal performance when only the first- and second-order statistics of the one-hop neighbors are available at each node. When no such statistics are available, the proposed linear message-passing in \eqref{eq:delta_sumprod} optimized by the blind offline adaptation scheme in \cite{Abdi20} obtains a near-optimal detection performance. 

In Fig. \ref{fig:PdPf1}, we see that both of the sum-product and max-product algorithms exhibit different performance levels for different values of $\zeta$ and their performance is bounded from above by the detection rate of the proposed linear message-passing algorithm. Moreover, Fig. \ref{fig:PdPf1} indicates that the values we choose for $\zeta$ or $c_0$ heavily affect the performance of the message-passing algorithms concerned. For instance, the detection rates of the max-product algorithm and EGC drop below that of the local sensing method for $\zeta = 1$ and $c_0 = 1$. We have observed in our simulations that the performance of the max-product algorithm is improved when the learning factor $\zeta$ is increased from 0.01 to 0.1 and then is degraded severely by a further increase in $\zeta$, indicating that an optimization of the message-passing iteration is needed. Optimal values for the parameters of the message-passing iteration are found by the proposed optimization method while the analysis provided in this paper justifies why the resulting detector outperforms the max-product algorithm. Note that, it is not clear how to determine the optimal value for $c_0$ when EGC is used. The proposed optimization framework solves this problem effectively since ECG is, in fact, a special case of  linear data-fusion. We see in Fig. \ref{fig:PdPf1} that the optimal performance of EGC is achieved by the proposed method while no information is available a priori. 

It is worth noting that, the existing works do not clarify how to best determine $J_{kj}$'s in the max-product operation. The analysis proposed in this paper clarifies that such an optimization is equivalent to designing an optimal linear fusion scheme. By comparing the detection rate of the max-product algorithm against that of the optimal linear fusion, the performance gain obtained by the proposed framework is visualized. Specifically, by comparing the curves labeled ''mp0.1'', ''mp0.3'', and ''mp1.0'' in Fig. \ref{fig:PdPf1} against the detection rates labeled ''linPropB'', ''linProp'', and ''linOpt'' we see the performance gain obtained. 

\section{Conclusion} \label{sec:conclusion}
The analysis and numerical results presented in this paper demonstrate that in a distributed detection scenario and under practical assumptions the max-product algorithm works as a linear data-fusion scheme. Therefore, the knowledge already developed in the literature regarding distributed linear data-fusion can be used to better understand the behavior of the max-product algorithm.

\bibliographystyle{IEEEtran}
\bibliography{IEEEabrv,Bibliogeraphy_ICLAS}

 \end{document}